\documentclass{PoS}
\usepackage{xcolor}
\graphicspath{{images/}}

\title{Semileptonic B decay results from early Belle II data}

\ShortTitle{Semileptonic B decay results from early Belle II data}

\author{\speaker{Andrea Fodor}\thanks{On behalf of the Belle II collaboration.}\\
        McGill University\\
        E-mail: \email{afodor@physics.mcgill.ca}}

\abstract{The Belle II experiment at the SuperKEKB energy-asymmetric $e^+ e^-$ collider is a substantial upgrade of the B factory facility, Belle, at the Japanese KEK laboratory.
The main operation of SuperKEKB has started in March 2019, and collisions ran until July 2019, achieving a peak luminosity of $5.5\times 10^{33}$ cm$^{-2}$s$^{-1}$. 
The results presented here were obtained from a subset of collected data of 0.41~fb$^{-1}$.
In the poster presented at the conference the first results from studying semileptonic $\textit{B}$-meson decays were shown. The performance of the Full Event Interpretation (FEI) tagging was analysed. The $B^0 \rightarrow D^{*-} \ell^+ \nu_{\ell}$ mode was rediscovered, using the untagged approach. 
}

\FullConference{XXIX International Symposium on Lepton Photon Interactions at High Energies - LeptonPhoton2019\\
		August 5-10, 2019\\
		Toronto, Canada}

\begin{document}

\section{Introduction}
The Belle II experiment\cite{b2book} is a B-factory experiment at the SuperKEKB $e^+e^-$ collider. The design luminosity of the machine is $8\times 10^{35}$ cm$^{-2}$s$^{-1}$ and the Belle II experiment aims to record 50 ab$^{-1}$ of data, a factor of 50 more than its predecessor. The first physics run of the Belle II experiment with the full Belle II detector took place from March until July 2019, recording 6.49~fb$^{-1}$ of data. This presentation shows the results from the first 0.41~fb$^{-1}$ of data, the first properly calibrated dataset ready in time for LP conference. For more details about the first data-taking results, see the talk by T. Browder at this conference\cite{tb}. 
\section{$\textit{B}$-meson reconstruction and Full Event Interpretation}
Electrons and positrons are collided at the $\Upsilon(4S)$ resonance, which decays almost exclusively to $\textit{B}$-meson pairs. $\textit{B}$ mesons are produced almost at rest in the center-of-mass frame. Reconstruction of the signal $\textit{B}$ meson from a desired decay mode, $B_{sig}$, as well as the reconstruction of the accompanying  $\textit{B}$ meson, $B_{tag}$, enables us to infer the event kinematics and account for any missing energy. There are several approaches in the analysis depending on the treatment of the $B_{tag}$.\\
In untagged reconstruction, $B_{sig}$ is reconstructed without full $B_{tag}$ reconstruction. It has high efficiency and high background contribution. The tagged approach involves reconstruction of the $B_{tag}$ using semileptonic or hadronic decay modes and attributing the remaining detected depositions to the $B_{sig}$. This approach lowers the background contributions, provides better control of the event kinematics, but also has lower efficiency compared to an untagged reconstruction.
Full Event Interpretation (FEI)\cite{fei} is a novel tagging approach implemented by Belle II that uses machine learning and reconstructs $B_{tag}$ from more than 200 different decay modes and over 10 000 decay chains, improving the $B_{tag}$ reconstruction efficiency compared to previous approaches. It enables precise determination of the four momentum of undetected particles. FEI outputs a classifier value, $\mathcal{P}_{tag}$, which discriminates between correctly reconstructed tag-side decay modes and combinatorial and physics backgrounds.
\section{Hadronic FEI tagging performance}
\begin{figure}[h]
\centering
\includegraphics[width=0.55\textwidth]{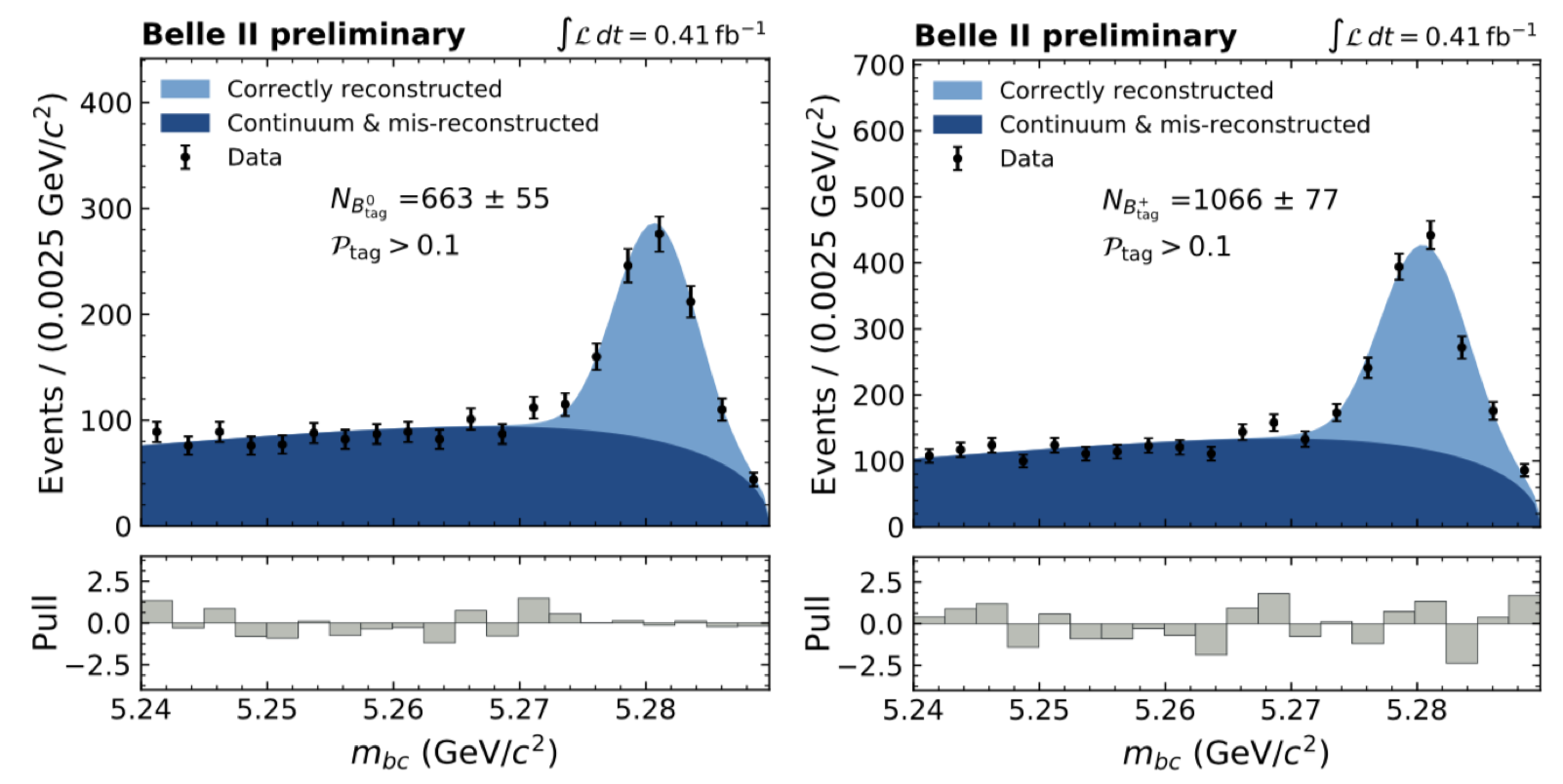}
\vspace{-0.3cm}
\caption{Fits to the beam-constrained mass, $m_{bc}$: distribution for reconstructed $B^0$ (left) and $B^+$ (right) tag-side $\textit{B}$ mesons in data, with a cut on the FEI classifier, $\mathcal{P}_{tag}>0.1$. Correctly reconstructed signal is modelled with a Cystal Ball and mis-reconstructed $\textit{B}$ mesons and continuum are modelled with an Argus shape.}
\label{fei_performance}
\end{figure}

\begin{figure}[h]
\centering
\includegraphics[width=0.58\textwidth]{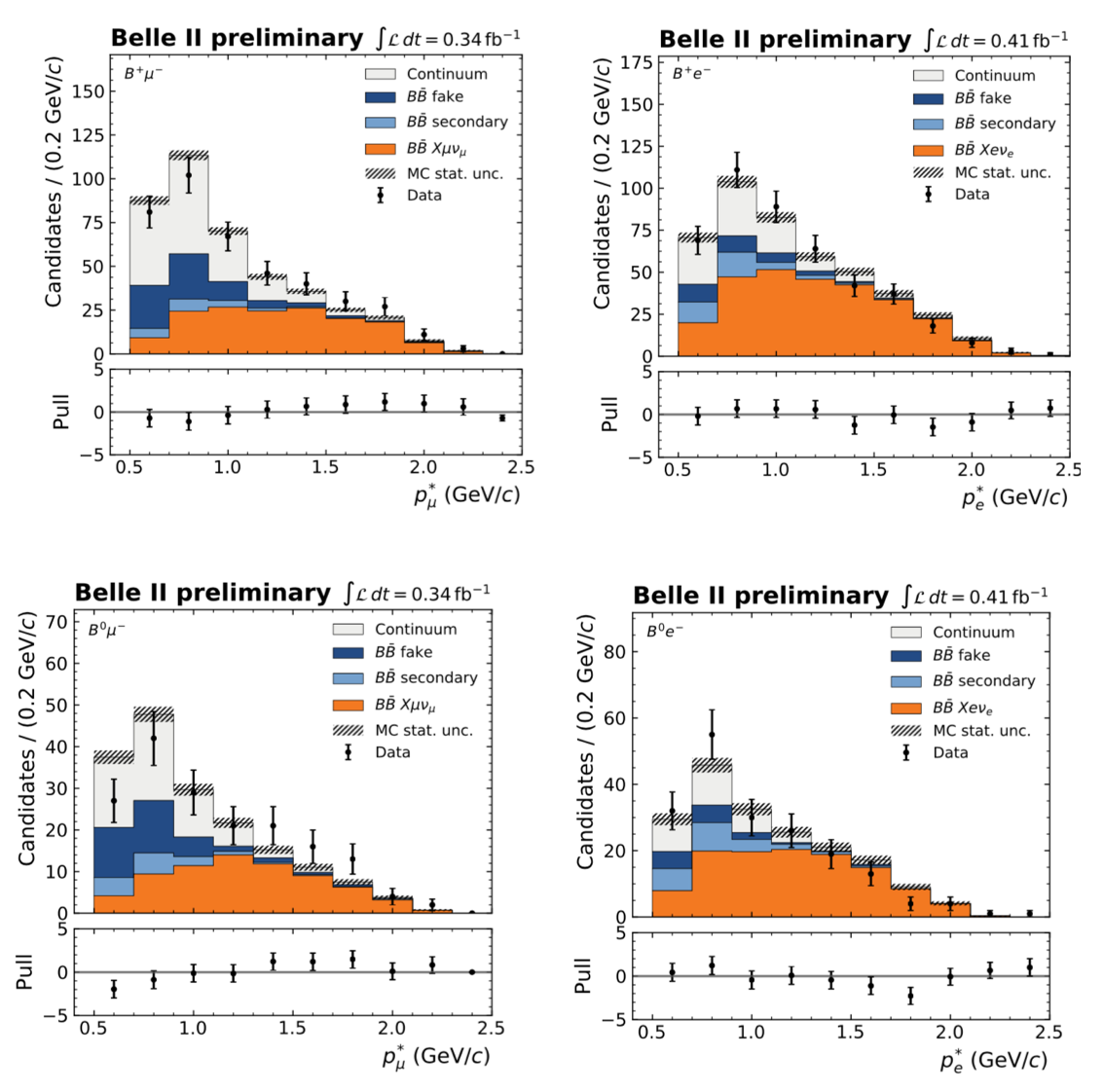}
\vspace{-0.3cm}
\caption{Distribution of the center-of-mass momentum for muons, $p_{\mu}^*$ (left), and electrons, $p_e^*$ (right), where the companion $\textit{B}$ meson, $B_{tag}$, is charged (top) or neutral (bottom).}
\label{xlnu_had}
\end{figure}

The first data collected by the Belle II experiment was analyzed to validate the FEI performance. For reconstruction, 29 and 26 hadronic $B^+$ and $B^0$ tag-side decay modes were used, respectively. The distribution of beam-constrained mass, $m_{bc}$, of the $B_{tag}$ is shown in Figure \ref{fei_performance}, where 
$$m_{bc} = \sqrt{s/4 - |\vec{p}^{*}_{B_{tag}}|^2},$$ 
where $s$ is the center-of-mass energy of the colliding $e^+e^-$-pair, and $\vec{p}^{*}_{B_{tag}}$ is the center-of-mass momentum of the reconstructed $B_{tag}$. A cut on the FEI classifier output of $\mathcal{P}_{tag}>0.1$ was applied.
Further, the hadronic FEI performance was examined using the $B \rightarrow X \ell \nu_{\ell}$ decay mode. The highest momentum charged lepton was selected from the tracks not associated with $B_{tag}$. The selected lepton is required to have the center-of-mass momentum $p^*_l>0.6$~GeV/c. Comparisons between data and Monte Carlo simulation for muon and electron candidate momenta are shown in Figure~\ref{xlnu_had}.\\
\section{$B^0 \rightarrow D^{*-} \ell^+ \nu_{\ell}$ - untagged reconstruction}
\begin{figure}[h]
\centering
\includegraphics[width=0.58\textwidth]{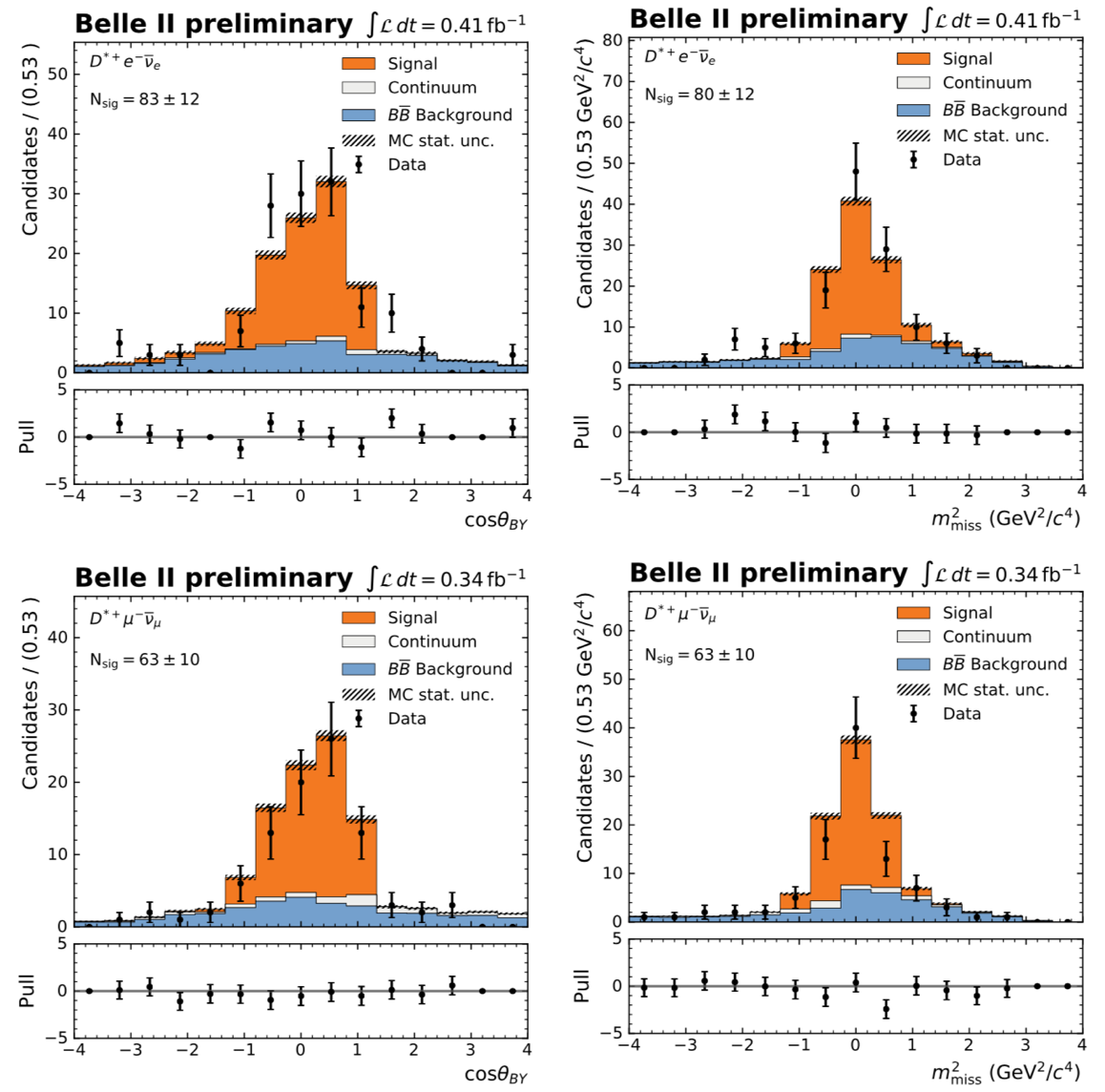}
\vspace{-0.3cm}
\caption{Distributions of $\cos\theta_{BY}$(right) and $m^2_{miss}$(left) for electron(top) and muon(bottom) modes for $B^0 \rightarrow D^{*-} \ell^+ \nu_{\ell}$ decay mode are shown.}
\label{dlnu}
\end{figure}

Rediscovery of the decay $B^0 \rightarrow D^{*-} \ell^+ \nu_{\ell}$ is an excellent way to validate the detector performance and software reconstruction, given its large branching fraction of $(4.95\pm0.11)\%$ \cite{bfdlnu}.
This mode will be used for $|V_{cb}|$ and form-factor measurements in the future. $D^{*\pm}$ candidates were reconstructed from the $D^{*\pm} \rightarrow D^{0} \pi^{\pm}$ decays. $D^0$ mesons were reconstructed in the $K^{-} \pi^{+}$ decay mode. Electron or muon candidates were combined with the $D^{*\pm}$ candidates to form signal candidates. Signal events can be identified using either the missing mass squared or the angle between the $\textit{B}$ meson and the $D^* \ell$ system:
$$m^2_{miss} = (\frac{p_{ee}}{2} - p_Y)^2, \qquad \cos\theta_{BY} = \frac{2E^*_B E^*_Y - M^2_B - m^2_Y}{2 p^*_B p^*_Y},$$
where $E_Y^*$, $p^*_Y,$ and $m_Y$ are the CM energy, momentum, and invariant mass of the $D^*\ell$ system, $M_B$ is the nominal B mass, and $E_B^*$, $p^*_B$ are the CM energy and momentum of the B, inferred from the CM machine energy.
The distributions of $m^2_{miss}$ and $\cos\theta_{BY}$ are shown in Figure~\ref{dlnu}. A total of $(80\pm12)$ and $(63\pm10)$ events were observed in the electron and muon mode, respectively, as determined with a binned likelihood fit. The yields are in good agreement with the expectations from the Monte Carlo simulations.
\section{Summary}
The first 0.41~fb$^{-1}$ of data from the Belle II experiment were analysed to validate the detector and software performance and rediscover missing energy decays. The FEI has been shown to be able to reconstruct hadronic $B_{tag}$ modes, and further studies of the signal-side with $B \rightarrow X \ell \nu_{\ell}$ have been carried out.
 Further studies 
on the performance were done using the $B^0 \rightarrow D^{*-} \ell^+ \nu_{\ell}$ mode, leading to the rediscovery of this decay mode. The successful first run is a gateway to fruitful results from the semileptonic and missing energy B-decays from Belle II in the future. 


\begin{thebibliography}{99}
\vspace{-0.25cm}
\bibitem{tb}T.~Browder, \emph{\textit{Recent News from Belle II}}, to appear in PoS(LeptonPhoton2019) 
\bibitem{b2book}E.~Kou, P.~Urquijo et al. (Belle II collaboration) 10.1093/ptep/ptz106, arXiv:1808.10567 [hep-ex].
\bibitem{fei}T.~Keck, F.~Abudinén, F.U.~Bernlochner et al. Comput Softw Big Sci (2019) 3: 6.
\bibitem{bfdlnu}M.~Tanabashi et al. (Particle Data Group), Phys. Rev. D \emph{\bf{98}}, 030001 (2018) and 2019 update. 
\end{thebibliography}
\end{document}